# On the specifics of the electrical conductivity anomalies in PVC nanocomposites

D.V.Vlasov, L.A.Apresyan

ABSTRACT

A qualitative model describing the "anomalous" features of the conductivity of polymer nanocomposites, in particular, switching to the conducting state in relatively thick (tens of microns or more) of flexible PVC films is considered. In previously published experimental results, change of conductivity by 10 or more orders of magnitude occurred both in the absence of external influences (spontaneously), and under the influence of an applied electric field, as well as other initiating factors (such as uniaxial pressure) . In a model of hopping conduction mechanism it is shown, that switching in the conduction states under the action of external field significantly (by orders of magnitude) below threshold can be associated with a high-resistance state instability that results from the sequence of "shorting" (reversible soft breakdown) of narrow insulating gaps between regions with relatively high conductivity. Increasing the field strength in the remaining insulating gaps ultimately leads to the formation of a conducting channel between the external electrodes and switching conductivity of the composite film sample in a state of high conductivity. This cascade model is essentially based on the transition from the usual description of the charge tunneling through single independent insulating gap to take into account correlations between adjacent gaps. In the frame of developed model other "anomalies" such as exponential dependence of the resistance on the sample thickness, pressure, and other influences can be qualitative explained. An analogy of the model with a cascading breakdown of avalanche transistors is also considered.

.

1.Введение.

В ряде работ авторов [1-5] были обнаружены и исследовались аномальные переключения электропроводности в прозрачных пленках пластикатов ПВХ с антистатическим уровнем электропроводности без использования гетерогенных проводящих наполнителей при толщине пленок от десятков до сотен микрон. В недавних работах [6,7] аналогичные аномальные эффекты были обнаружены авторами также в ПВХ-пленках, в которых для повышения проводимости использовалось создание двойных сопряженных связей путем термического дегидрохлорирования ПВХ в растворе. В таких пленках при напряженностях поля значительно меньших уровня порога пробоя наблюдаются спонтанные и стимулированные внешним воздействием обратимые переходы из исходного антистатического состояния в состояние с высокой проводимостью (СВП). Отметим, что скачок проводимости составлял 4 порядка и более для пластифицированных пленок, и до 12 порядков величины в случае термолизованного ПВХ.

В тонких пленках нанометрового диапазона скачки проводимости возникают практически в любых диэлектриках при приближении к пороговому значению напряженности поля («мягкий пробой»), и, в частности, широко исследовались в пленочных системах металл-диэлектрик-металл, в том числе в пленках полимеров [8-11].
В противоположность и в дополнение к этим специфическим «тонкопленочным» явлениям в наших исследованиях [1-7] спонтанные и стимулированные скачки проводимости обнаруживались в относительно толстых пленках, до сотен микрон и более. Для многомикронных пленок широкозонных полимеров при относительно низких напряжениях порядка вольт такое поведение можно назвать аномальным, так как во многих других исследованиях [9-11] специально отмечалось существование граничной критической толщины $L_{кр}$ при превышении которой, скачки проводимости не наблюдались, причем обычно наблюдаемая толщина $L_{кр}$ ограничивается субмикронным или даже нанометровым диапазоном [11].

Поскольку переключения проводимости в полимерных пленках представляют большой прикладной интерес, в частности, в связи с возможностями создания новых элементов памяти [12,13], в литературе большое внимание уделяется попыткам объяснения физических механизмов наблюдаемых переключений. Однако, при всем многообразии и специфике исследуемых материалов, разнообразии предложенных механизмов достаточно много экспериментальных результатов не нашло своего объяснения и, как отмечается авторитетными исследователями, окончательное решение вопроса о наиболее вероятных механизмах таких переключений все еще не получено (см., напр., обзоры [11-13]).

В данной заметке мы предлагаем для интерпретации аномалий каскадный механизм применительно к композитной среде, когда в образце преобладает прыжковая проводимость. При этом серия «мягких» микропробоев реализует каскадный механизм развития неустойчивостей, что позволяет качественно объяснить основные характеристики наблюдаемых ранее в [1-7] спонтанных и стимулированных переключений между двумя состояниями проводимости в относительно толстых, порядка десятков микрон пленках, при уровнях напряжений на порядки ниже порогов «мягкого» пробоя. При этом в рамках развиваемой модели нет необходимости конкретизировать микроскопическую теорию физических механизмов «мягкого» пробоя, которые даже в случае тонких нанометровых полимерных пленок могут иметь разнообразный характер, зависящий от материала полимера и конкретных структур полимерных пленок, и приводить к разным вольт-амперным характеристикам, наблюдаемым в экспериментах [13]. Можно ожидать, что рассматриваемый простой механизм может оказаться полезным для понимания других экспериментальных результатов по измерению электропроводности, в первую очередь, в случае композитных диэлектрических пленок, содержащих проводящие включения.

2. Экспериментальные факты, стимулирующие построение модели.

Поведение электропроводности пластифицированных и дегидрохлорированных пленок ПВХ экспериментально исследовалось нами в цитированных работах [1-7]. Здесь мы приведем лишь некоторые экспериментальные результаты, которые стимулировали развитие и были использованы при построении модели переходов в СВП.

Прежде всего, в последних экспериментах [7] с сополимером ПВХ-полиацетилен, при переходе в СВП сопротивление образца толщиной 30 мкм составляло менее десятой Ома. Для другой геометрии измерений того же самого образца при эффективной толщине измерительного зазора 3 мм в схеме измерения поверхностного сопротивления, в соответствие с формулой для сопротивления образца R через его геометрию (L-длина, S-площадь поперечного сечения) и удельное сопротивление (ρ):

$$R=\rho L/S, \qquad (1)$$

связанное с объемными токами поверхностное сопротивление должно было увеличиться всего лишь в $10^6$ раз и прекрасно измеряться прибором с пределом $10^{12}$ Ом. Тем не менее несмотря на многочисленные попытки, замерить поверхностное сопротивление образца так и не удалось, поскольку оно превышало $10^{12}$ Ом. Таким образом, для образца сополимера ПВХ-полиацетилен повсеместно используемая формула (1) давала ошибку более чем на шесть порядков, что конечно требовало привлечения дополнительных моделей для интерпретации. Ранее и в пластифицированной пленке при измерении зависимости от толщины образца была получена экспоненциальная зависимость от толщины, что очевидным образом противоречит общепринятой формуле расчета (1).

Основной механизм, привлекаемый для интерпретации аномальных данных, связан с результатами, полученными в работе [2] для ПВХ пленок, где с повышением приложенного пилообразного напряжения переходы в СВП происходят при достижении порогового напряжения $E_{пор}$ (порог «мягкого» пробоя), сопровождающиеся сбросами в исходное состояние в виде некоторого стохастического процесса, отвечающего обратимым случайным переходам в СВП и в исходное низкопроводящее состояние (см. Рис.1). При этом, как отмечалось в [2], переключения сопровождались генерацией отчетливо различимых звуковых волн, и кроме того в той же работе аналогичный «мягкий» пробой наблюдался в других пластиковых пленках, полиэтилена, тефлона и др. В стандартной схеме измерений при наличии большого балластного сопротивления такие переходы происходят обратимо, без разрушения полимерной пленки и материалов электродов.

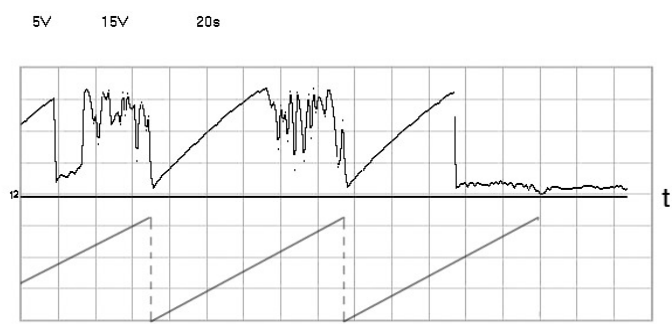

Рис.1.Последовательные осциллограммы напряжения на образце при циклах сканирования приложенного к измерительной ячейке напряжения; в нижней части рисунков указана пилообразная форма импульсов подаваемого напряжения [2].

Из Рис.1. видно, что при достижении некоторого порогового напряжения $E_{пор}$ (Вольт) достаточно регулярно происходит «мягкий» пробой, обратимость которого доказывается многократностью переключений. В частности, в этом случае после относительно быстрого переключения проводимости вследствие мягкого пробоя образец некоторое время $\tau_м$ находится в проводящем состоянии, в том числе и при полном снятии внешнего напряжения. В частности, как отмечалось в работе [2], образец сутками мог оставаться в проводящем состоянии с последующим возвратом в нормальное состояние с низким уровнем проводимости. Кроме того, распределение проводимости по пленке было крайне неоднородно [2], и собственно, реализовалось в нескольких точках на поверхности образца (проводящих каналах, аналогичных наблюдаемым в других работах, см., напр., [14]).

Динамика однократных спонтанного и стимулированного переключения образцов пластикатов ПВХ в СВП проиллюстрирована в работах [3,5]. На Рис.2 (по материалам работ [7,8]) показана осциллограмма однократного спонтанного переход в СВП в образце дегидрохлорированного ПВХ. Из Рис.2 следует, что за относительно быстрым переключением в СВП следует некоторый интервал времени $\tau_м$ , в примере на Рис.2 - несколько секунд нахождения образца в высоко проводящем состоянии, а затем - возвращение в исходное состояние, которое происходит более плавно, причем полностью исходная проводимость восстанавливается лишь через десятки секунд, что позволяет оценить порядок величины $\tau_м$ .

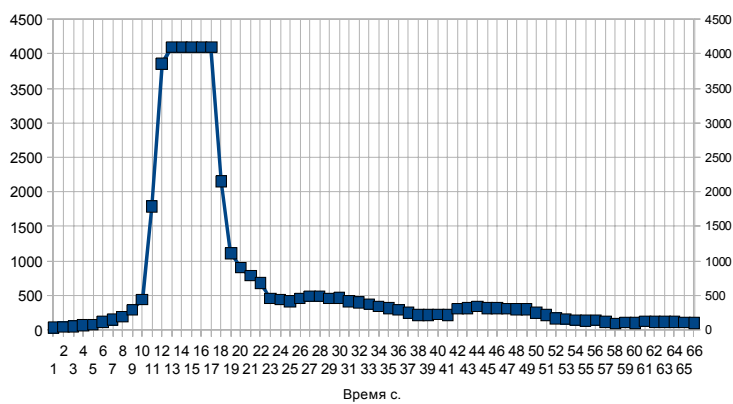

Рис. 2. Спонтанный переход в СВП в пленочном образце ПВХ-полиацетилен нанокомпозите.

Таким образом композиты ПВХ, как и многие другие полимеры и диэлектрики-изоляторы, имеют четко выраженный устойчивый порог «мягкого» пробоя или иными словами обратимого, неразрушающего перехода в СВП, причем возврат из СВП происходит не мгновенно при снятии поля, а может характеризоваться временем $\tau_р$, зависящим от многих параметров.

3. Каскадная модель развития переходов в СВП

В случае термического дегидрохлорирования ПВХ из-за отщепления хлористого водорода возникает сополимер ПВХ-полиацетилен, т.е. в макромолекулы ПВХ встраиваются цепочки сопряженных двойных связей, которые обладают повышенной электропроводностью. Величина этой электропроводности может приближаться к металлической, если присутствующие в ПВХ атомы хлора выступают в качестве допантов (как известно, именно допирование позволяет существенно - на 7-10 порядков - поднять проводимость сопряженных полимеров, см.,напр., [15] ). Учитывая это, для описания электропроводности образцов дегидрохлорированного ПВХ в указанных выше относительно толстых пленках примем модель композита, состоящего из хорошо проводящих областей, отделенных друг от друга слоями диэлектрика (исходного чистого ПВХ).

Для вероятности туннелирования электрона под потенциальным барьером $U(x)$ приближение ВКБ дает хорошо известное выражение [16]:

$$W = \exp(-(\sqrt{8m}/\hbar) \int \sqrt{U(x)-\varepsilon}\, dx), \qquad (2)$$

где $U(x)= U_0 - qEx$ потенциал барьера, а q-заряд электрона, $\hbar$ – постоянная Планка, а m и $\epsilon$ - соответственно, масса и энергия электрона, причем в нашем случае $\epsilon \sim kT \ll U_0$ . Отсюда видно, что зависимость прозрачности барьера от его толщины в основном экспоненциальна, но зависит также от внешнего поля, причем возрастает экспоненциально при его увеличении. Так, в простейшем приближении $U_0 >> \epsilon + qEx$ для прямоугольного барьера шириной $\Delta x$ выражение (2) для вероятности туннелирования приводит к выражению

$$W(\Delta x, U_0) = \exp(-\Delta x / l_d), \qquad (3)$$

где $1/l_d = \sqrt{8mU_0} / \hbar$ – характерная длина, зависящая от высоты барьера $U_0$. Отметим, что соотношение (2) по теореме о среднем можно использовать и в общем случае, но при этом характерная длина $l_d$ будет сложным образом зависеть от формы барьера и напряженности поля.

В предлагаемой модели вероятность туннелирования через 2 соседних барьера будет определяться просто произведением вероятностей (3) (поскольку события независимы). При этом вероятность снова определяется экспонентой, причем показатели экспонент (3) суммируются. При расчете полной вероятности перескока электрона с одного электрода на другой в рамках сделанных выше предположений получаем полную длину потенциального барьера равную сумме всех длин, $L_в = \Sigma \delta x_i$. Таким образом, полная вероятность туннелирования с электрода А на электрод Б выражается экспонентой, причем в показатель входит суммарная ширина всех потенциальных барьеров. Анализируя все возможные траектории движения заряда от А к Б, можно ожидать, что для большинства траекторий туннелирования электрона вероятность экспоненциально мала, тем не менее из множества случайных траекторий существует траектория с некоторым минимальным значением $L_в$. Причем очевидно, что реально имеет шанс «сработать» только траектория с минимальным показателем экспоненты (3). Вдоль такой траектории от А к Б можем записать

$$L = L_в + L_с, \qquad (4)$$

где L – расстояние между электродами, а $L_с$ - полная длина проводников. Тогда среднюю напряженность поля, с учетом эквипотенциальности проводящих элементов можно оценить как:

$$E = U_{АБ} / L_в = U_{АБ} / (L - L_с). \qquad (5)$$

Это простое выражение показывает, что при уменьшении полной длины барьера, локальные напряженности поля на барьерах возрастает. В частности, для теории перколяции [17] из этой простой формулы вытекает, что по мере увеличения коцентрации проводящей примеси и, соответственно, приближения к образованию бесконечного кластера, локальная напряженность поля может значительно возрастать и система (до порога перколяции) за счет туннелирования заряда будет переходить в проводящее состояние и может совершать переходы аналогичные приведенным на рис.1.

Наоборот, при снижении концентрации проводящей фазы вероятность «мягкого пробоя» будет экспоненциально затухать с ростом толщины изолирующего зазора $\Delta x$ при фиксированной разности потенциалов $\Delta \varphi \sim E \Delta x$.

Рассмотрим теперь более детально модельную цепочку, упорядоченную вдоль оптимальной трассы туннелирования вдоль приложенного однородного внешнего поля, и состоящую из проводящих областей, изолированных друг от друга для простоты одинаковыми зазорами $\Delta x$ (Рис.3). Напряженность поля E в каждом из зазоров при оценках можно считать одинаковой, так что разности потенциалов $\varphi_i$ проводников 1,2 и 3 равны, $\Delta \varphi = \varphi_2 - \varphi_1 = \varphi_3 - \varphi_2$. В случае возникновения «мягкого пробоя» зазора 12 потенциалы проводников 1 и 2 выравниваются (на некоторое время $\tau_м$), а разность потенциалов в зазоре 23 удваивается, что приводит к росту вероятности пробоя этого зазора. В свою очередь, при пробое зазора 23 увеличивается разность потенциалов между областью 3 и соседней справа от нее областью, находящейся при большем потенциале, что приводит к увеличению вероятности следующего пробоя и т.д.

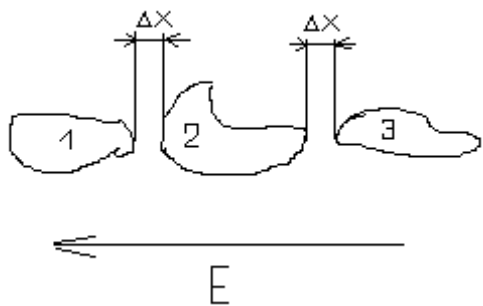

Рис.3. Модель цепочки из проводящих областей с изолирующими зазорами.

Таким образом, с учетом порога мягкого пробоя и конечного времени жизни проводящего микро- канала проводимости мягкого пробоя, можно говорить о каскадном механизме развития неустойчивости вдоль приложенного внешнего поля или вдоль наиболее вероятной траектории туннелирования (канала).

Напротив, если в некотором промежутке канал мягкого пробоя восстанавливает изоляционные свойства, то поле соседних интервалов уменьшается, что соответственно увеличивает вероятность возврата в исходное состояние. Таким образом, вполне спонтанно могут развиваться неустойчивости, как по возникновению СВП, так и по возврату в исходное состояние. Подчеркнем, что в простейшем варианте модели (одинаковой высоты потенциальных барьеров) она предсказывает экспоненциальную зависимость от суммарной толщины изолирующих промежутков, т.е. при фиксированной концентрации проводящей фазы от толщины образца.

4. О механизме спонтанного перехода в СВП («свитчинга»).

Как известно, даже чистый полимер имеет сложное (фрактальное) распределение плотности, большое количество специфических дефектов, кристаллитов и т.д. при наличии свободного объема, во многом определяющего физические свойства полимеров. При помещении полимера или полимерного композита в однородное внешнее электрическое поле в объеме среды могут возникать сильные флуктуации напряженности, которые могут существенно превышать величину среднего значения приложенного поля, и достигать значений, при которых может происходить «мягкий» обратимый пробой тонких слоев изоляторов.

Самую грубую оценку таких флуктуаций можно получить, рассмотрев макроскопическую модель скругленного острия с радиусом r на расстоянии d от проводящей поверхности, что дает фактор усиления ~ $(d/r)^2$. Если принять в качестве d среднее расстояние между проводящими областями, рассматривая r как характерный размер «выступа» на границе проводящей области, то уже при r ~ d/3 (что для сильно-неоднородных сред представляется вполне допустимым) получается усиление на порядок по сравнению со средним значением поля.

Отметим, что значительно большие коэффициенты усиления (порядка $10^5$ и более) получаются в моделях так называемых метаматериалов - композитов металл-диэлектрик при наличии плазмонных резонансов [18], однако в рассматриваемом случае трудно ожидать наличия таких коллективных эффектов, связанных с отрицательными значениями эффективных диэлектрических проницаемостей металлических областей.

Наряду с флуктуациями поля разумно предположить, что в пластичной полимерной среде могут флуктуировать изолирующие зазоры (в особенности даже при слабых механических вибрациях), соответственно еще сильнее флуктуируют и локальные поля, в частности, с превышением порога мягкого пробоя. В таком случае переход в СВП начинается с «мягкого пробоя» одного из тонких зазоров между проводящими областями, в которых возникают свободные заряды или реализуется наибольшие локальные значения напряженности электрического поля (при этом мы оставляем в стороне вопрос о деталях микроскопического механизма такого пробоя). Инициирующим фактором такого пробоя в случае тонких зазоров естественно считать туннелирование электрона через зазор.

С другой стороны, стимулировать переключение могут и космические частицы, ионизирующие и создающие свободные заряды и облегчающие мягкий пробой изолирующего промежутка и развития неустойчивости. Аналогично, на переключение в проводящее состояние может влиять освещенность и появление фотоэлектронов в изолирующем промежутке, микро-смещения при наложении на образец внешнего давления и другие воздействия. Экспериментальные результаты, подтверждающие такие переходы, были получены как в наших экспериментах [1-8], так и во многих других [9].

5. Область применимости модели.

Таким образом, каскадная модель мягкого пробоя на качественном уровне полностью позволяет объяснить наблюдавшиеся в композитной среде (сополимер ПВХ-полиацетилен) аномальные переключения проводимости в относительно толстых пленках и абсолютную неприменимость формулы (1) для расчета сопротивления образца через объемное удельное сопротивление (поскольку сопротивление экспоненциально зависит от толщины образца).

Более того, несмотря на то, что модель сформулирована для модельной композитной среды типа проводник-диэлектрик, что представляется наиболее обоснованным в случае дегидрохлорированного ПВХ, содержащего цепочки с двойными связями, многие результаты, которые она объясняет и предсказывает, применимы и для описания аномальных эффектов пластикатов пленок ПВХ, исследованных ранее [1-5]. Это может свидетельствовать о наличии в таких пластикатах областей повышенной проводимости. В частности, в таких пластикатах наблюдались спонтанные и стимулированные переходы в СВП, а также переходы в СВП под действием давления. На рис. 4 приведена измеренная в [5] экспоненциальная зависимость сопротивления образца от его толщины.

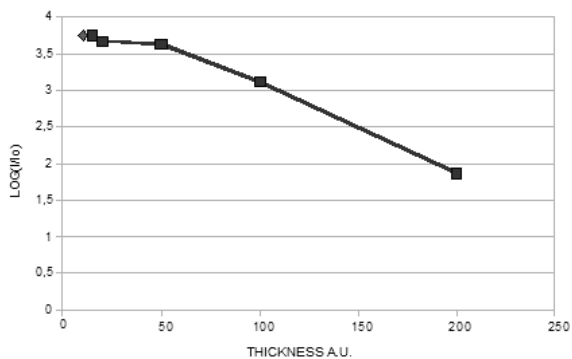

Рис.4. Экспериментальная зависимость тока от толщины образца пластиката при фиксированном напряжении. По оси абсцисс толщина образца в произвольных единицах близких к мкм,- по оси ординат десятичный логарифм тока [5].

Экспоненциальная зависимость сопротивления от толщины образца полностью объясняет произвольное отклонение от общепринятой формулы расчета сопротивления (1) в случае перехода в СВП, и неприменимости в этом случае к полимерам (и композитам на базе полимеров) собственно величины удельного сопротивления.

В качестве еще одного примера неадекватности (1) применительно к полимерным материалам можно привести результаты работы [9], в которой наблюдалась более сложная экспоненциальная зависимость измеряемого сопротивления от толщины образца.

6.Об аналогах модели

Ранее, в работе [2] было отмечено сходство мягкого пробоя полимерных пленок с известным лавинным пробоем полупроводниковых триодов, который также наступает при определенном пороговом $E_{пор}$, многократно обратим при достаточно большом сопротивлении нагрузки и имеет характерное время жизни $\tau$, существенно превышающее время развития собственно пробоя. Лавинный пробой часто используется в электронике для генерации коротких (обычно порядка долей наносекунды) фронтов. При этом, если требуется получить импульс большой амплитуды, много больше $E_{пор}$. то транзисторы включают последовательно, причем N последовательно включенных транзисторов дают порог пробоя $N \cdot E_{пор}$ (при этом для выравнивания напряжений используется резисторный делитель напряжения). В такой схеме также реализуется каскадный механизм пробоя. полностью аналогичный описанному выше: при повышении напряжения в одном из транзисторов реализуется лавинный пробой, при этом напряжение на других резко повышается и они практически мгновенно переходят в проводящее состояние обеспечивая на выходе мощный импульс тока и напряжения. Существенное отличие в рассмотренной выше модели полимерной среды состоит в том, что аналог «транзисторов» в полимерной среде, т.е. зазоры между проводящими областями случайны, поэтому переходы в СВП могут происходить при различных напряжениях, в том числе спонтанно, поскольку наиболее короткие промежутки-изоляторы могут переходить в СВП при очень малых относительно $E_{пор}$ напряжениях. Таким образом, приготовленные специальным образом полимер-композитные пленки могут уже сегодня использоваться в схемах каскадных генераторов импульсов вместо лавинных транзисторов.

Отметим также, что проявление нелинейности даже в слабых внешних полях, сопровождающееся резким изменением сопротивления, наблюдаемое в полимерных композитах, а также наличие времени запаздывания $\tau_м$, которое может быть значительным, указывают на возможность создания на базе рассматриваемых полимерных композитых сред новых пассивных элементов — мемристоров.

7.Выводы.

Рассмотрена прыжковая модель проводимости в полимер-композитных материалах, которая с учетом экспериментально установленного порога мягкого пробоя (обратимого перехода в состояния высокой проводимости) и конечного времени жизни канала мягкого пробоя позволила объяснить на качественном уровне некоторые зависимости и механизмы переключений проводимости ПВХ полимер-композитов, которые ранее считались аномальными.

В частности, такая модель объясняет спонтанные и стимулированные различными внешними воздействиями переключения проводимости, возврат проводимости в исходное состояние, экспоненциальную зависимость сопротивления образца от его толщины. Мы полагаем, что рассмотренная модель при ее дальнейшем развитии позволит получить корректные оценки, а в ряде случаев и количественно оценить поведение электропроводности и для других полимерных нанокомпозитов.

ЛИТЕРАТУРА